\documentstyle{article}
\title {
\vspace*{-2.1cm}\hspace*{8.7cm} {\large gr-qc/yymm,
CGPG-94/4-4}\\
\bf \vspace*{1.2cm} Bases of Wormholes in Quantum Cosmology}
\author{\vspace*{.5cm} \\Guillermo A. Mena Marug\'an\\ \\
Center for Gravitational Physics and Geometry,\\ Pennsylvania State
University, University Park, PA 16802, USA.\\
\\ On leave from: Instituto de Matem\'aticas y F\'{\i}sica Fundamental,\\
C.S.I.C., Serrano 121, 28006 Madrid, Spain.\vspace*{.5cm} \\}
\date{March, 1994}

\begin{document}
\renewcommand{\thefootnote}{\fnsymbol{footnote}}

\maketitle
\large
\setlength{\baselineskip}{.825cm}
\begin{center}
{\Large {\bf Abstract}}
\end{center}

\vspace*{.4cm}

We show that if the space of physical states spanned by the wormhole
wave functions can be equipped with a Hilbert
structure, such a Hilbert space must coincide with that of the Lorentzian
gravitational system under consideration. The physical inner product can then
be determined by imposing a set of Lorentzian reality conditions. The
Hilbert space of the gravitational model admits in this case a basis of
wormhole solutions, and every proper quantum state can be interpreted as a
superposition of wormholes. We also argue that the
wave functions that form the basis of wormholes must be eigenfunctions of a
complete set of compatible observables. The associated eigenvalues provide
a set of well-defined wormhole parameters, in the sense that they can
be employed to designate the different elements of the basis of wormholes.
We analyse in detail the case of a Friedmann-Robertson-Walker spacetime
minimally coupled to a massless scalar field. For this minisuperspace, we
prove the validity of all the above statements and discuss various admissible
choices of bases of wormhole wave functions.

\newpage

\section {Introduction}

Wormholes are solutions to the equations of motion of Euclidean General
Relativity that connect two asymptotic three-manifolds of large volume [1,2].
They can be interpreted as closed universes that branch off, or join onto,
the asymptotic flat regions of spacetime [1]. It has been argued that the
existence of wormholes may significantly affect the effective low-energy
interactions, altering the value of the physical constants [1] and providing
a mechanism for the vanishing of the cosmological constant [3].

If wormholes
are to play a crucial role in physics, they should occur regardless of the
specific properties of the matter content of the Universe. However, classical
solutions of the wormhole type can exist only for very special kinds of matter
fields, namely, those which allow the Ricci tensor to have
negative eigenvalues [4]. None the less, if it is possible to construct a
quantum theory of gravity, wormholes do not need to be realized as classical
solutions to the Einstein equations [5]. Quantum mechanically, a wormhole
will simply be a physical state (wave function) with a characteristic
asymptotic behaviour in the limit of large three-geometries, such that it can
represent a tube connected to a region of spacetime with large three-volume
[5].
These considerations led Hawking and Page to propose that the wormhole
wave functions should be exponentially damped for large three-geometries, and
regular for degenerate ones [5]. The last requirement guarantees that the
wave functions present no singularities when the three-geometry collapses
to zero because of an ill-defined slicing of the spacetime. These wormhole
boundary conditions have also been reformulated in a language that is
specially well-suited for the Euclidean path integral approach to Quantum
Cosmology. According to Garay [6], the wormhole wave functions can thus be
represented as path integrals over Euclidean manifolds that match asymptotic
large three-geometries with no gravitational excitations, with the
gravitational momenta fixed in that asymptotic region. In particular,
the wormhole ground state is obtained by demanding that, in addition, there
exist no matter excitations in the regions of large volume.

In order to determine the effect of wormholes in the low-energy physics, it
has been assumed in the literature that the vector space spanned by the
wormhole wave functions can be equipped with a Hilbert structure [1,3].
Although this conjecture has never been proved in full gravity, its validity
has been vindicated at least in a collection of minisuperspace models in
which the complete set of wormhole states is known [5-8]. In all these
minisuperspaces, nevertheless, the inner product in the space of wormholes
has been constructed by restricting to the space of physical states
an inner product that is naturally defined in the unconstrained gravitational
models [7]. This proposal to fix the inner product must not be expected
to work satisfactorily in more general cases, because the physical states of
a constrained system need not be normalizable with respect to any pre-Hilbert
structure introduced before imposing the first-class constraints of the
theory. It therefore seems necessary to adopt a different prescription
to determine a unique inner product in the space of wormhole wave functions.

In the last years, Ashtekar has developed a detailed programme to achieve the
non-perturbative quantization of systems with first-class constraints,
including the determination of the physical inner product [9-11].
One must first select a (possibly over-)complete set of complex functions on
phase space that is closed under Poisson brackets, and promote it to an
abstract $\ast$-algebra of operators, with the classical Poisson brackets
translated into commutators and the complex conjugation relations between
classical variables captured in the $\ast$-involution [9,10]. This
$\ast$-algebra is represented on a chosen vector space, with the
physical states provided by the kernel of all the operators that represent the
first-class constraints of the theory. Finally, one must find a sufficiently
large number of observables, i.e., operators that commute with all the
constraints. The physical inner product is then uniquely fixed [12] by
imposing the $\ast$-relations between observables as adjointness conditions
[9,10]. These requirements are usually called reality conditions [9,13].

We can apply this systematic programme to those minisuperspaces models
for which the wormhole quantum solutions have been explicitly found, and
discuss in this way whether it is possible to arrive at a well-defined
Hilbert structure for the space of wormholes. Nevertheless,
let us notice first that the quantization approach proposed by
Ashtekar determines completely the Hilbert space of physical states once
one has specified a set of reality conditions and chosen a particular
representation (what includes in turn the selection of an algebra of
elementary operators). Suppose now that, for a given representation, the
space of wormholes admits a Hilbert structure, selected by a certain set of
reality conditions. We are then led to conclude that either the reality
conditions that correspond to the wormhole states are not Lorentzian, or the
wormhole wave functions must span a subspace of the physical (Lorentzian)
Hilbert space of the analysed quantum theory.

At first glance, one might think that, instead of being Lorentzian, the
reality conditions for wormholes should be those of Euclidean gravity,
because the wormhole wave functions can be represented
(at least in some models) as Euclidean path integrals [6]. However,
the Euclidean action that provides the weight of the different paths in those
integrals can be related to the Lorentzian gravitational action by means of a
Wick rotation, and we know that reality conditions are left invariant under
such rotations [14]. Hence, we should actually expect that the reality
conditions for the wormhole states are those corresponding to the Lorentzian
theory. Note in this sense that the wormhole wave functions satisfy in fact
the Lorentzian Wheeler-DeWitt equation, and not its Euclidean counterpart
(i.e., the equation obtained from the Euclidean Hamiltonian constraint
assuming the same commutation relations for Euclidean as for Lorentzian
gravity [14]). On the other hand, in those minisuperspaces in which the
wormhole solutions have been shown to form a Hilbert space (and even if we
do not agree with the prescription adopted to select the Hilbert structure),
the inner product was found by demanding that the Lorentzian momenta of
the gravitational and matter fields were self-adjoint [7], requirements that
can clearly be interpreted as Lorentzian reality conditions for the
corresponding unconstrained gravitational models. We thus claim that
the Hilbert structure of the space of wormholes must be determined
by imposing an appropriate set of Lorentzian reality conditions.

As a consequence, we conclude that if for a given representation there exists
a Hilbert space of wormholes, it must be a subspace of the Hilbert space of
the Lorentzian quantum theory. Moreover, if we consider only irreducible
representations (and assume that the space of wormholes is stable under the
action of the observables of the system), the wormhole wave functions have to
span the whole Hilbert space of physical states in the constructed
quantization. This means that the wormhole solutions provide in this case a
basis for the Hilbert space of the Lorentzian theory, and that the so-called
Hilbert space of wormholes [5,7] is just the physical Hilbert space of the
analysed gravitational model. Notice that, in general, the wormhole
wave functions that form a particular basis for the Hilbert space need not
be normalizable with respect to the physical inner product, although it may
be possible to find a change to a new basis of quantum states
that have finite norm. In any of these bases, the different wave functions
will be designated by a set of wormhole
parameters [1,3] which, in order to characterize the wormhole states, must
possess a well-defined meaning in the constrained gravitational theory
and must therefore correspond to eigenvalues of observables of the system.

In the rest of this work, we will show with an explicit example that the
Hilbert space of wormholes is nothing else but the Hilbert space of the
Lorentzian quantum theory. We will study the case of a
Friedmann-Robertson-Walker model in the presence of a minimally coupled
massless scalar field. For this minisuperspace, the complete set of wormhole
states is known [5]. In section 2 we introduce the model and carry out its
non-perturbative canonical quantization according to Ashtekar's programme,
obtaining in this way the physical inner product. We then find in section
3 the integral transform that leads from the representation used in this
non-perturbative quantization to the geometrodynamic representation in which
the wormhole wave functions were originally found. We also prove there that
the Hilbert space of the model is spanned by a basis of wormholes, and
discuss the normalizability properties of the wormhole states which form that
basis. In section 4 we analyse in detail other admissible choices for the
basis of wormholes. In particular, we show that there exists a discrete basis
of wormhole wave functions that are normalizable with respect to the
Lorentzian inner product. Finally, we present our conclusions and some
further remarks in section 5.

\section {Minisuperspace Model: Quantization}

We will consider a Friedmann-Robertson-Walker spacetime, described by the
metric
\begin{equation}
ds^2=\frac{2G}{3\pi}\left(-\frac{N^2(t)}{2}q^3(t)dt^2+2q(t)d\Omega_3^2\right)
,\;\;\;\;\;\;q(t)>0.\end{equation}
Here, $q(t)$ is the square of the scale factor of the model, $N(t)$ is the
rescaled lapse function and $d\Omega_3^2$ is the metric of the unit
three-sphere. The matter content of the system will consist of a minimally
coupled massless scalar field, $\Phi$, that will be assumed to be
homogeneous. Rescaling the scalar field as
\begin{equation} \Phi(t)=\sqrt{\frac{3}{16\pi G}}\phi(t)\end{equation}
the Hamiltonian constraint of the model can be expressed [5,6]:
\begin{equation} H\equiv \frac{1}{2}(q^2\Pi_q^2+q^2-\Pi_{\phi}^2)=0,
\;\;\;\;\;\;q>0,\end{equation}
with $\Pi_q$ and $\Pi_{\phi}$ the momenta canonically conjugate to $q$ and
$\phi$, respectively.

Let us introduce now the following canonical transformation of variables [15]:
\begin{equation} q=\frac{\Pi}{\cosh X},\;\;\; q\Pi_q=-\Pi \tanh
X.\end{equation}
A generating function for this transformation is provided by
\begin{equation} F(q,X)=-q \sinh X.\end{equation}
The first equation in (4) implies that the momentum $\Pi$
conjugate to $X$ is strictly positive, because $q>0$. The configuration
variable $X$, on the other hand, runs over the whole real axis, $X\in
I\!\!\!\,R$. Employing relations (4), the Hamiltonian constraint (3)
can be rewritten as
\begin{equation} H\equiv\frac{1}{2}(\Pi^2-\Pi_{\phi}^2)=0,\;\;\;\;\;\;\Pi>0.
\end{equation}
This is the only constraint of the system, since we have already fixed the
diffeomorphism gauge freedom.

Equation (6) is satisfied only if $\Pi_{\phi}=\Pi>0$ or $\Pi_{\phi}=
-\Pi<0$. As a consequence, the reduced phase space of the model splits into
two disconnected pieces, i.e., the cotangent bundles over the positive
($\Pi_{\phi}>0$) and the negative ($\Pi_{\phi}<0$) real axis. In fact, both
sectors ($\Pi_{\phi}>0$ and $\Pi_{\phi}<0$) are related by a time reversal
operation, so that they can be considered as physically equivalent [15].
Owing to this fact, we will restrict our attention in the following to the
case $\Pi_{\phi}>0$, unless explicitly stated.

To construct the quantum theory, let us choose as representation space the
space of complex functions $\Psi(X,\Pi_{\phi})$ on $I\!\!\!\,R\times
I\!\!\!\,R^{+}$, with  $X\in I\!\!\!\,R$ and $\Pi_{\phi}\in I\!\!\!\,R^{+}$.
In this representation (and taking $\hbar=1$ from now on), we can introduce
the following canonical set of basic operators:
\begin{equation} \hat{\Pi}_{\phi}\Psi=\Pi_{\phi}
\Psi(X,\Pi_{\phi}),\;\;\;\hat{\phi}\Psi=i\partial_{\Pi_{\phi}}
\Psi(X,\Pi_{\phi}),\end{equation}
\begin{equation} \hat{X}\Psi=X
\Psi(X,\Pi_{\phi}),\;\;\;\;\;\hat{\Pi}\Psi=-i\partial_{X}
\Psi(X,\Pi_{\phi}).\;\end{equation}
A simple calculation shows then that the physical states that satisfy
\begin{equation} \hat{\Pi}\Psi(X,\Pi_{\phi})=\Pi_{\phi}\Psi(X,\Pi_{\phi}),
\end{equation}
and thus the quantum version of the Hamiltonian constraint (6), adopt the
generic expression
\begin{equation}
\Psi(X,\Pi_{\phi})=f(\Pi_{\phi})\,e^{i\Pi_{\phi}X},\end{equation}
$f(\Pi_{\phi})$ being any complex function on $I\!\!\!\,R^{+}$.

A complete set of classical observables on the reduced phase space of the
model (the cotangent bundle over $I\!\!\!\,R^{+}$) is provided by $\Pi_{\phi}$
and [15]
\begin{equation} w=\Pi_{\phi}\phi+\Pi_{\phi}^2\frac{X}{\Pi}.\end{equation}
This set is closed under Poisson brackets, for
$\{w,\Pi_{\phi}\}=\Pi_{\phi}$. In the quantum theory, the action of these
observables on the physical states (10) will be given by the first equation
in (7) and
\begin{equation} \hat{w}\left(f(\Pi_{\phi})e^{i\Pi_{\phi}X}\right)=
i\Pi_{\phi}\left(\partial_{\Pi_{\phi}}f(\Pi_{\phi})\right)
e^{i\Pi_{\phi}X}.\end{equation}
This formula can be obtained from (7-9) and the definition of $w$ (11)
with an adequate choice of factor ordering.

For Lorentzian gravity, the classical variables $\Pi_{\phi}$ and $w$ are real.
Therefore, the reality conditions associated with the Lorentzian theory
demand the hermiticity of
the operators $\hat{\Pi}_{\phi}$ and $\hat{w}$ on the space of
physical states. These requirements select the inner product
\begin{equation}
<\Theta,\Psi>=\int_{I\!\!\!\,R^+}\frac{d\Pi_{\phi}}{\Pi_{\phi}}
g^{\ast}(\Pi_{\phi}) f(\Pi_{\phi}),\end{equation}
where $\Theta(X,\Pi_{\phi})=g(\Pi_{\phi}) e^{i\Pi_{\phi}X}$ and the symbol
$^{\ast}$ denotes complex conjugation. The Hilbert space of physical states is
thus the space of wave functions of the form (10) with $f(\Pi_{\phi})\in
L^2(I\!\!\!\,R^+,d\Pi_{\phi}/\Pi_{\phi})$. On the other hand, it is not
difficult to check that the obtained representation is irreducible.

Similar conclusions could have been reached if we had analysed instead the
sector of negative momenta $\Pi_{\phi}$. In that case, one can choose as
representation space the space of complex functions $\Psi(X,\Pi_{\phi})$ on
$I\!\!\!\,R\times I\!\!\!\,R^-$, with $\Pi_{\phi}<0$. The physical states are
then the wave functions
\begin{equation}
\Psi(X,\Pi_{\phi})=\tilde{f}(\Pi_{\phi})\,e^{-i\Pi_{\phi}X},\end{equation}
with $\tilde{f}(\Pi_{\phi})$ a generic function on $ I\!\!\!\,R^-$. They
satisfy the constraint
\begin{equation} \hat{\Pi}\Psi(X,\Pi_{\phi})=-\Pi_{\phi}\Psi(X,\Pi_{\phi}).
\end{equation}
The Lorentzian reality conditions determine now the inner product
\begin{equation}
<\Theta,\Psi>=\int_{I\!\!\!\,R^-}\frac{d\Pi_{\phi}}{|\Pi_{\phi}|}
\tilde{g}^{\ast}(\Pi_{\phi}) \tilde{f}(\Pi_{\phi}),\end{equation}
where $\Theta(X,\Pi_{\phi})=\tilde{g}(\Pi_{\phi}) e^{-i\Pi_{\phi}X}$.
Thus, the Hilbert space consists of all functions (14) with
$\tilde{f}(\Pi_{\phi})\in L^2(I\!\!\!\,R^-,d\Pi_{\phi}/|\Pi_{\phi}|)$.
It is then easy to show that the representations of the quantum theory
constructed here for $\Pi_{\phi}>0$ and $\Pi_{\phi}<0$ are unitarily
equivalent.

In the next section, we will translate these results into the geometrodynamic
representation in which the wormhole wave functions of this model have been
obtained [5-7]. In order to do that, it will be useful to introduce
first a different set of elementary operators in the $(X,\Pi_{\phi})$
representation that we have studied. The reason for this is that the action of
these new elementary operators will have a direct counterpart in the
geometrodynamic representation that we are going to analyse. Let us then
define (either for  $\Pi_{\phi}>0$ or $\Pi_{\phi}<0$) the operators
\begin{equation} \hat{\Pi}_{\phi}\Psi=\Pi_{\phi}
\Psi(X,\Pi_{\phi}),\;\;\widehat{\Pi_{\phi}\phi}\Psi=
i\left(\Pi_{\phi}\partial_{\Pi_{\phi}}-\frac{1}{2}\right)\Psi(X,\Pi_{\phi}),
\end{equation}
\begin{equation} \widehat{(\Pi\tanh X)}\Psi=\!-i\partial_X\!\left(\tanh X
\Psi(X,\Pi_{\phi})\right),\;\,\widehat{\left(\frac{\Pi}{\cosh X}\right)}\Psi
=\!-i\partial_X\!\left(\frac{\Psi(X,\Pi_{\phi})}{\cosh X}\right),
\end{equation}
with only non-vanishing commutators given by
\begin{equation} [ \widehat{\Pi_{\phi}\phi},\hat{\Pi}_{\phi} ]=
i\hat{\Pi}_{\phi},\;\;\; [\widehat{(\Pi \tanh X)},\widehat{\left(\frac{\Pi}
{\cosh X}\right)}]=i
\widehat{\left(\frac{\Pi}{\cosh X}\right)},\end{equation}
so that they form a closed algebra under commutation relations.
These operators represent a complete set of elementary variables, as can be
proved by employing relations (4) and recalling that, for $q>0$ and
$\Pi_{\phi}>0$ (or $\Pi_{\phi}<0$), the phase space of the unconstrained
theory
is the cotangent bundle over $(I\!\!\!\,R^+)^2$ (or $I\!\!\!\,R^+\times
I\!\!\!\,R^-$, respectively). Equation (17) ensures that
the action of $\widehat{\Pi_{\phi}\phi}$ on the physical
states (10) or (14) coincides with that derived from relation (11)
(and (7-9) and (15)) with the following factor ordering
\begin{equation} \widehat{\Pi_{\phi}\phi}=\hat{w}-\hat{\Pi}_{\phi}^2
\frac{1}{2}(\hat{X}\hat{\Pi}^{-1}+\hat{\Pi}^{-1}\hat{X}).\end{equation}
The definitions (18), on the other hand,
can be obtained from equation (8) with a particular choice of operator
ordering,
i.e., with the derivative operator $\hat{\Pi}$ acting on the left of any
factor depending on $X$. This ordering guarantees that
the operator $\hat{\Pi}^2$ acts exactly as the sum of the squares
of $\widehat{(\Pi\tanh X)}$ and $\widehat{(\frac{\Pi}{\cosh X})}$.
Hence, the quantum version
of the Hamiltonian constraint (6) can be rewritten as
\begin{equation} \hat{H}\Psi(X,\Pi_{\phi})\equiv\frac{1}{2}\left[
\widehat{(\Pi\tanh X)}^2+\widehat{\left(\frac{\Pi}{\cosh
X}\right)}^2-\hat{\Pi}_{\phi}^2\right]\Psi(X,\Pi_{\phi})=0.\end{equation}

\section {Change of Representation}

We are now in an adequate position to discuss the relation between the
$(X,\Pi_{\phi})$ representation, used to perform the non-perturbative
quantization, and the geometrodynamic $(q,\phi)$
representation in which the wormhole wave functions of the model are
known [5-7]. Before carrying out any explicit calculations, let us notice that
in the $(q,\phi)$ representation studied in the literature [5,7] the momenta
$\Pi_{\phi}$ has been assumed to run over the whole real axis,
instead of being restricted to have either positive or negative values.
The situation is in fact similar to that which one encounters when analysing
the quantum theory of a relativistic particle subject to a Klein-Gordon
constraint. Such theory admits two equivalent irreducible
representations, one for positive frequencies and the other for negatives
ones. In the position representation, however, the wave functions are
usually allowed to get contributions from both types of frequencies.
The representation
obtained is not longer irreducible\footnote{Assuming that we do not allow
time reversal operations}, but given by the direct sum of two irreducible
representations, that correspond, respectively, to the sectors of positive and
negative frequencies. Extrapolating these results to our model, we expect the
geometrodynamic
$(q,\phi)$ representation to be provided by the direct sum of the irreducible
representations constructed for $\Pi_{\phi}>0$ and
$\Pi_{\phi}<0$ in section 2. Adding these representations, the physical
states of the theory adopt the generic expression
\begin{equation} \Psi(X,\Pi_{\phi})=\left\{\begin{array}{cr}f(\Pi_{\phi})
e^{i\Pi_{\phi}X} & \;\;\Pi_{\phi}>0\\ \tilde{f}(\Pi_{\phi})
e^{-i\Pi_{\phi}X} & \;\;\Pi_{\phi}<0\end{array}\right.\;\;,\end{equation}
and the physical inner product can be taken as
\begin{equation}
<\Theta,\Psi>=\int_{I\!\!\!\,R^+}\frac{d\Pi_{\phi}}{\Pi_{\phi}}
g^{\ast}(\Pi_{\phi})f(\Pi_{\phi})+\int_{I\!\!\!\,R^-}\frac{d\Pi_{\phi}}
{|\Pi_{\phi}|}\tilde{g}^{\ast}(\Pi_{\phi})\tilde{f}(\Pi_{\phi}),\end{equation}
where $(g(\Pi_{\phi}),\tilde{g}(\Pi_{\phi}))$ are the functions that
characterize the state $\Theta(X,\Pi_{\phi})$. Note that, in defining this
inner
product, there actually exists an ambiguity of a positive factor which
determines the relative weight of the contributions from the
representations with positive and negative momenta $\Pi_{\phi}$.
Since these representations are physically
equivalent, we have chosen that factor to be equal to the unity.

The change to the $(q,\phi)$ representation must be given by an integral
transform of the type [16]
\[ \Psi(q,\phi)=\sigma(\Psi(X,\Pi_{\phi}))=\int_{I\!\!\!\,R^+}d\Pi_{\phi}
g_1^+(\Pi_{\phi})e^{i\phi\Pi_{\phi}}\int_{I\!\!\!\,R}dX g_2^+(X)
e^{iF(q,X)}\,f(\Pi_{\phi})e^{i\Pi_{\phi}X}+ \]
\begin{equation}\int_{I\!\!\!\,R^-}d\Pi_{\phi}
g_1^-(\Pi_{\phi})e^{i\phi\Pi_{\phi}}\int_{I\!\!\!\,R}dX g_2^-(X)
e^{iF(q,X)}\,\tilde{f}(\Pi_{\phi})e^{-i\Pi_{\phi}X}.\end{equation}
Here, $\sigma$ denotes the change of representation, and $F(q,X)$ is the
generating function (5). The integral transform in $X$ must provide the
change to some intermediate $(q,\Pi_{\phi})$ representation, both for
$\Pi_{\phi}>0$ and $\Pi_{\phi}<0$. The generalized Fourier transform in
$\Pi_{\phi}$ leads then to desired $(q,\phi)$ representation,
obtained as the direct sum of two irreducible pieces.
The functions
$(g_1^+,g_2^+,g_1^-,g_2^-)$ that appear in formula (24) can be fixed by
requiring that, under the integral transform, the operators (17,18),
defined in the $(X,\Pi_{\phi})$ representation, act on $\Psi(q,\phi)$
in the following way:
\begin{equation} \sigma(\hat{\Pi}_{\phi}\Psi(X,\Pi_{\phi}))\equiv
\hat{\Pi}_{\phi}\Psi(q,\phi)=-i\partial_{\phi}\Psi(q,\phi),\end{equation}
\begin{equation}\sigma\left(\widehat{\left(\frac{\Pi}{\cosh
X}\right)}\Psi(X,\Pi_{\phi})\right)\equiv
\hat{q}\Psi(q,\phi)=q\Psi(q,\phi),\end{equation}
\begin{equation}
\sigma(-\widehat{(\Pi\tanh X)}\Psi(X,\Pi_{\phi}))\equiv
\widehat{q\Pi_q}\Psi(q,\phi)=-iq\partial_q\Psi(q,\phi),\end{equation}
\begin{equation}
\sigma(\widehat{\Pi_{\phi}\phi}\Psi(X,\Pi_{\phi}))\equiv
\widehat{\Pi_{\phi}\phi}\Psi(q,\phi)=-i\left(\phi\partial_{\phi}+\frac{1}{2}
\right)\Psi(q,\phi).\end{equation}
In equations (26,27) we have used relations (4) to define the quantum
operators $\hat{q}$ and $\widehat{q\Pi_q}$. Equations (25,26), on the other
hand, reproduce the standard action of the operators $\hat{\Pi}$ and $\hat{q}$
in the $(q,\phi)$ representation. Condition (27) guarantees then that, under
the change of representation, the quantum Hamiltonian constraint (21)
translates into
\begin{equation} \hat{H}\Psi(q,\phi)\equiv \frac{1}{2}\left(q^2
-(q\partial_q)^2+\partial_{\phi}^2\right)\Psi(q,\phi)=0,\end{equation}
which is exactly the Wheeler-DeWitt equation (associated with constraint
(3)) that is solved by the known wormhole wave functions of the model
[5-7]. Thus, equation (27), together with (25,26), leads precisely to the
factor ordering that has been employed in the literature for the
geometrodynamic analysis of this minisuperspace.
Finally, requirement (28) allows us to interpret the action of
$\widehat{\Pi_{\phi}\phi}$ in the $(q,\phi)$ representation as the
symmetrized product of the multiplicative operator $\hat{\phi}$ (with
$\hat{\phi}\Psi=\phi\Psi(q,\phi)$) and $\hat{\Pi}_{\phi}=-i\partial_{\phi}$.

A careful calculation shows then that, for all those physical states that
are normalizable with respect to the inner product (23), the change of
representation (24) is well-defined and satisfies conditions (25-28)
if and only if $g_2^+$ and $g_2^-$ are non-vanishing constants and
$g_1^+$ and $g_1^-$ are proportional to $\Pi_{\phi}^{-1}$. So, we can let
in the following
\begin{equation} g_1^+=g_1^-=\frac{1}{\Pi_{\phi}} ,\;\;\;\;g_2^+=g_2^-=1.
\end{equation}
Using now that, for $q>0$ [17],
\begin{equation} \int_{I\!\!\!\,R}dXe^{-iq\sinh{X}+i\Pi_{\phi}X}=
2e^{\frac{\pi}{2}\Pi_{\phi}}K_{i\Pi_{\phi}}(q),\end{equation}
with $K_{i\Pi_{\phi}}$ a modified Bessel function of imaginary order,
equation (24) can be rewritten in the compact form
\begin{equation} \Psi(q,\phi)=\int_{I\!\!\!\,R^+}\frac{d\Pi_{\phi}}
{\Pi_{\phi}}2e^{\frac{\pi}{2}\Pi_{\phi}}K_{i\Pi_{\phi}}(q)\left(
e^{i\phi\Pi_{\phi}}f(\Pi_{\phi})+e^{-i\phi\Pi_{\phi}}h(\Pi_{\phi})\right),
\end{equation}
where we have introduced the notation
\begin{equation} h(\Pi_{\phi})= -\tilde{f}(-\Pi_{\phi}),\;\;\;\;\Pi_{\phi}
>0.\end{equation}
Hence, $h(\Pi_{\phi})\in L^2(I\!\!\!\,R^+,d\Pi_{\phi}/\Pi_{\phi})$
for states $\Psi$ in the physical Hilbert space, because in that case
$\tilde{f}(\Pi_{\phi})\in L^2(I\!\!\!\,R^-,d\Pi_{\phi}/|\Pi_{\phi}|)$.

The transformation (32) can be inverted to recover the expression of the
wave functions in the $(X,\Pi_{\phi})$ representation. The inversion is
given by the formulas
\begin{equation} f(\Pi_{\phi})=\int_{I\!\!\!\,R^+}dq G(\Pi_{\phi},q)
\int_{I\!\!\!\,R}d\phi e^{-i\phi\Pi_{\phi}}\Psi(q,\phi),\end{equation}
\begin{equation}h(\Pi_{\phi})=
\int_{I\!\!\!\,R^+}dq G(\Pi_{\phi},q)\int_{I\!\!\!\,R}d\phi
e^{i\phi\Pi_{\phi}}\Psi(q,\phi),\end{equation}
\begin{equation} G(\Pi_{\phi},q)=\frac{2\Pi_{\phi}}
{\pi^2}e^{-\frac{\pi}{2}\Pi_{\phi}}\cosh{(\pi\Pi_{\phi})}K_{i\Pi_{\phi}}(q).
\end{equation}
In deriving these equations, we have employed that [18]
\begin{equation} \int_0^{\infty}dq \,K_{i\Pi_{\phi}}^2(q)=\frac{\pi^2}
{4\cosh(\pi\Pi_{\phi})}\;.\end{equation}
Substituting then (34-36) in (23), it is possible to derive the explicit
form of the physical inner product in the geometrodynamic representation
$(q,\phi)$.

Equation (32) implies in particular that, in the adopted $(q,\phi)$
representation, the Hilbert space of physical states is spanned by the
functions
\begin{equation} \Psi_{\epsilon\tilde{\Pi}_{\phi}}(q,\phi)=\frac{2}{\sqrt{
\tilde{\Pi}_{\phi}}}e^{\frac{\pi}{2}\tilde{\Pi}_{\phi}}K_{i\tilde{\Pi}_{\phi}}
(q)e^{i\epsilon\phi\tilde{\Pi}_{\phi}},\end{equation}
where $\epsilon=1$ or $-1$ and $\tilde{\Pi}_{\phi}$ is any positive constant
($\tilde{\Pi}_{\phi}\in I\!\!\!\,R^+$). These functions are physical states
of the model [5,7], and can be obtained from the generic formula (32)
by letting
\begin{equation} f(\Pi_{\phi})=\sqrt{\Pi_{\phi}}\delta(\Pi_{\phi}-\tilde{
\Pi}_{\phi})\equiv f_{+\tilde{\Pi}_{\phi}}(\Pi_{\phi}),\;\;h(\Pi_{\phi})=0
\equiv h_{+\tilde{\Pi}_{\phi}}(\Pi_{\phi})\end{equation}
if $\epsilon=1$, or
\begin{equation}f(\Pi_{\phi})=0\equiv f_{-\tilde{\Pi}_{\phi}}(\Pi_{\phi}),\;\;
h(\Pi_{\phi})=\sqrt{\Pi_{\phi}}\delta(\Pi_{\phi}-\tilde{
\Pi}_{\phi})\equiv h_{-\tilde{\Pi}_{\phi}}(\Pi_{\phi})\end{equation}
when $\epsilon=-1$. Recalling equation (33), it is then straightforward to
check that these wave functions are not normalizable with respect to the
inner product (23). Nevertheless, they satisfy the orthogonality conditions
\begin{equation} <\Psi_{\epsilon'\tilde{\Pi}_{\phi}'},\Psi_{\epsilon
\tilde{\Pi}_{\phi}}>=\delta_{\epsilon'\epsilon}\delta(\tilde{\Pi}_{\phi}'-
\tilde{\Pi}_{\phi}),\end{equation}
with $\delta_{\epsilon'\epsilon}$ the Kronecker delta. In this way, the set
of wave functions (38) turns out to provide an orthogonal (but not
normalizable) basis for the Hilbert space of the model in the $(q,\phi)$
representation.

The physical states (38) were first found by Hawking and Page when
discussing the existence of quantum wormhole solutions in the considered
minisuperspace [5]. To facilitate the comparison of our results with those
of [5,7], we note that the wave functions (38) with $\epsilon=-1$ can
be equivalently rewritten as
\begin{equation} \Psi_{-\tilde{\Pi}_{\phi}}(q,\phi)=\tilde{\Psi}_{\tilde{\Pi}
_{\phi}'}(q,\phi)=\frac{2}{\sqrt{|\tilde{\Pi}'_{\phi}|}}e^{\frac{\pi}{2}
|\tilde{\Pi}'_{\phi}|}K_{i\tilde{\Pi}_{\phi}'}(q)e^{i\phi\tilde{\Pi}_{\phi}'}
,\end{equation}
where $\tilde{\Pi}_{\phi}'=-\tilde{\Pi}_{\phi}<0$ and we have used that
$K_{i\tilde{\Pi}_{\phi}}(q)=K_{-i\tilde{\Pi}_{\phi}}(q)$. Actually, the
physical states (38) can be considered as generalized wormhole wave
functions, in the sense that, even though they fail to fulfil the
requirement of regularity when the three-geometry collapse to zero ($q
\rightarrow 0$), they are all exponentially damped for large scale factors
($q\rightarrow \infty$) [5]. Thus, neglecting for the moment the regularity
condition at $q=0$, we can assert that the space of states spanned in this
model by the quantum wormhole solutions does not only admit a Hilbert
structure, but coincides in addition with the Hilbert space of the Lorentzian
theory. Moreover, in the
next section we will show that, with a change of basis, it is possible to
find a complete set of states that are genuine wormhole solutions and
therefore do not display any singularity at the origin $q=0$.

The different wave functions (38) can be designated by the
number $\epsilon\tilde{\Pi}_{\phi}$, which is the eigenvalue
taken by the observable $\hat{\Pi}_{\phi}$ in these wormhole states.
In this way, the wormhole parameter of the model
($\epsilon\tilde{\Pi}_{\phi}$) turns out to be the eigenvalue of a quantum
observable, in agreement with our comments in the Introduction.

Finally, equation (32) allows us to express the physical states in the
$(q,\phi)$ representation as the following superposition of wormhole
wave functions:
\begin{equation}\Psi(q,\phi)=\int_{I\!\!\!\,R^+}d\Pi_{\phi}\left(\frac
{f(\Pi_{\phi})}{\sqrt{\Pi_{\phi}}}\Psi_{+\Pi_{\phi}}(q,\phi)+
\frac{h(\Pi_{\phi})}{\sqrt{\Pi_{\phi}}}\Psi_{-\Pi_{\phi}}(q,\phi)
\right).\end{equation}
The functions $f(\Pi_{\phi})$ and $h(\Pi_{\phi})$ that describe the quantum
states in the $(X,\Pi_{\phi})$ representation provide then, when
divided by $\sqrt{\Pi_{\phi}}$, the respective contributions of the wormhole
solutions $\Psi_{+\Pi_{\phi}}$ and $\Psi_{-\Pi_{\phi}}$ to the physical
state $\Psi(q,\phi)$.

\section {Bases of Normalizable Wormholes}

In the previous section we have proved that the Lorentzian reality conditions
select a well-defined inner product in the $(q,\phi)$ representation, and
that the Hilbert space determined by that inner product is spanned by a set
of generalized wormhole solutions. We want to show now that it is possible to
find other bases for the Hilbert space of the model such that their
elements represent proper wormhole states which are regular
everywhere in the configuration space $(q\in I\!\!\!\,R^+$, $\phi \in I\!\!\!
\,R$) and decrease exponentially for large scale factors $q\gg 1$.
In addition, the wormhole wave functions that form the bases to be
considered will be seen to be normalizable with respect to
the Lorentzian inner product (23).

Let us study first the set of wormhole solutions given by
\begin{equation} \Psi_{\phi_0}(q,\phi)=e^{-q\cosh{(\phi-\phi_0)}}
,\end{equation}
where $\phi_0\in I\!\!\!\,R$. These states can be obtained as path integrals
over asymptotically flat manifolds and over matter fields that take the value
$\phi_0$ in the asymptotic region of large three-volume [6]. These wave
functions are clearly regular when the three-geometry collapses to zero
($q\rightarrow 0$) and exponentially damped for $q\rightarrow \infty$.
Thus, they represent genuine wormhole solutions.

The physical states (44) can be recovered from equation (32) with the
choice
\begin{equation} f(\Pi_{\phi})=\frac{1}{2\pi}e^{-i\Pi_{\phi}\phi_0}
e^{-\frac{\pi}{2}\Pi_{\phi}}\Pi_{\phi}\equiv f_{\phi_0}(\Pi_{\phi}),
\end{equation}
\begin{equation} h(\Pi_{\phi})\equiv
h_{\phi_0}(\Pi_{\phi})=f_{\phi_0}^{\ast}(\Pi_{\phi}),
\end{equation}
as can be easily checked by taking into account that [18]
\begin{equation} \int_{I\!\!\!\,R}dp\, e^{ip(\phi-\phi_0)}K_{ip}(q)=\pi e^{-q
\cosh{(\phi-\phi_0)}},\;\;\;\;q>0.\end{equation}
{}From relation (47) it also follows that the generalized wormholes (38) can
be expressed as superpositions of the wave functions $ \Psi_{\phi_0}(q,\phi)$:
\begin{equation}\Psi_{\epsilon\tilde{\Pi}_{\phi}}(q,\phi)=\frac{2}
{\sqrt{\tilde{\Pi}_{\phi}}}e^{\frac{\pi}{2}\tilde{\Pi}_{\phi}}
\int_{I\!\!\!\,R}d(\phi-\phi_0)
e^{i\epsilon\tilde{\Pi}_{\phi}\phi_0} \Psi_{\phi_0}(q,\phi).\end{equation}
The set of wave functions $\Psi_{\epsilon\tilde{\Pi}_{\phi}}$ and
$\Psi_{\phi_0}$ result then in spanning the same space of physical
states. Therefore, the wormhole solutions (44) provide a basis
for the Hilbert space of the model in the $(q,\phi)$ representation.
Equation (48) can then be interpreted as the formula for the change of basis
from the set $\Psi_{\epsilon\tilde{\Pi}_{\phi}}$ ($\epsilon\tilde{\Pi}_{\phi}
\in I\!\!\!\,R$) to the new set   $\Psi_{\phi_0}$
($\phi_0\in I\!\!\!\,R$).

The wormhole wave functions (44) are completely characterized by the
value of the parameter
$\phi_0$. Using equations (22), (33) and (45,46) to describe these
wormhole states in the $(X,\Pi_{\phi})$ representation, it is not difficult
to check that the
wormhole parameter $\phi_0$ coincides in fact with the eigenvalue reached by
the observable
\begin{equation} \left( e^{-\frac{\pi}{2}|\hat{\Pi}_{\phi}|}\hat{w}
 e^{\frac{\pi}{2}|\hat{\Pi}_{\phi}|}\hat{\Pi}_{\phi}^{-1}\right)\end{equation}
in the considered wormhole solutions, with $\hat{\Pi}_{\phi}$ and $\hat{w}$
defined, respectively, by (7) and (11) (both for $\Pi_{\phi}>0$ and
$\Pi_{\phi}<0$).

Finally, from equations (33) and (45,46) it is straightforward to prove
that the wormhole wave functions (44) are normalizable with respect to the
inner product (23), so that they all represent proper physical states of
the quantum theory.

Besides the normalizable basis (44), there exists another known complete set
of wormhole solutions for the model that we are studying [5,7]. The elements
of this set adopt the expression
\begin{equation} \Psi_n(q,\phi)=\frac{1}{2^n n! \sqrt{\pi}}
H_n\left(\sqrt{2q}\cosh{\frac{\phi}{2}}\right)H_n\left(\sqrt{2q}
\sinh{\frac{\phi}{2}}\right)e^{-q\cosh{\phi}}.\end{equation}
Here, $n$ is a non-negative integer ($n=0,1,...$) and $H_n$ are the Hermite
polynomials [18]. These wave functions vanish exponentially when $q\rightarrow
\infty$, present no singularities in the region $\{q>0,\,\phi\in I\!\!\!\,R\}$
and possess a finite limit when $q$ tends to zero. In this sense, they can all
be considered as truly quantum wormholes. In addition, they form a discrete
basis for the Hilbert space of the system, since the wave functions (38) can
be written as the following combinations of the wormholes $\Psi_n$ [7]:
\begin{equation}\Psi_{\epsilon\tilde{\Pi}_{\phi}}(q,\phi)=\sum_{n=0}^{\infty}
F_n(\epsilon\tilde{\Pi}_{\phi})\Psi_n(q,\phi),\end{equation}
\begin{equation}
F_n(\epsilon\tilde{\Pi}_{\phi})=2\sqrt{\frac{\pi}{\tilde{\Pi}_{\phi}}}
e^{\frac{\pi}{2}\tilde{\Pi}_{\phi}}\int_{I\!\!\!\,R}d\eta \frac{\sinh^n\eta}
{\cosh^{n+1}\eta}e^{i2\eta\epsilon\tilde{\Pi}_{\phi}}.\end{equation}

The wormhole states (50) are eigenstates of the operators
\begin{equation}  \frac{1}{2}(x^2-\partial_x^2)\;\;\;{\rm and}\;\;\;
\frac{1}{2}(y^2-\partial_y^2)\end{equation}
with eigenvalues equal to $n+\frac{1}{2}$, where [5,7]
\begin{equation} x=\sqrt{2q}\cosh{\frac{\phi}{2}},\;\;\;
y=\sqrt{2q}\sinh{\frac{\phi}{2}}.\end{equation}
We then expect the operators (53) to be observables of the model,
for the integer $n$ corresponds to a well-defined wormhole parameter.
That this is indeed the case can be proved by realizing that,
from relations (54), one can rewrite the quantum Hamiltonian constraint
of the system (29) as
\begin{equation} \hat{H}\equiv 8(x^2-y^2)(x^2-y^2-\partial_x^2+\partial_y^2)=0
.\end{equation}
The operators (53) commute weakly with the above Hamiltonian, because,
for $q>0$, equations (54) imply that $x$ is always greater than $|y|$.

On the other hand, using formulas (34-36) one can find the explicit
expression of the wormhole wave functions (50) in the $(X,\Pi_{\phi})$
representation. A careful calculation leads to the result:
\begin{equation}
f(\Pi_{\phi})=\frac{1}{\pi\sqrt{\pi}}\cosh{(\pi\Pi_{\phi})}
e^{-\frac{\pi}{2}\Pi_{\phi}}\Pi_{\phi}\int_{I\!\!\!\,R}d\eta
\frac{\sinh^n\eta}{\cosh^{n+1}\eta}e^{-i2\eta\Pi_{\phi}}
\equiv f_n(\Pi_{\phi}),\end{equation}
\begin{equation} h(\Pi_{\phi})\equiv h_n(\Pi_{\phi})
=(-1)^n f_n(\Pi_{\phi}).\end{equation}
The wave functions $\Psi_n(X,\Pi_{\phi})$ can then be obtained from these
equations and relations (22) and (33). In particular, the wormhole states
(50) turn out to be normalizable with respect to the inner product (23),
because the functions $(f_n(\Pi_{\phi}),h_n(\Pi_{\phi}))$ are analytic for
all positive $\Pi_{\phi}$, vanish when $\Pi_{\phi}$ tends to zero and
decrease at infinity at least as fast as $\Pi_{\phi}^{-1}$, for it can
be proved that
\begin{equation} \left| \lim_{\Pi_{\phi}\rightarrow \infty}
e^{\frac{\pi}{2}\Pi_{\phi}}\Pi_{\phi}^2\int_{I\!\!\!\,R}d\eta
\frac{\sinh^n\eta}{\cosh^{n+1}\eta}e^{-i2\eta\Pi_{\phi}}\right|<\infty.
\end{equation}
Therefore, the wave functions (50) represent proper physical states of the
quantum theory. Finally, let us notice that the Hilbert space of the model is
separable, because the basis (50) is countable and normalizable.
As an additional consequence, the basis of wormholes $\Psi_n$
($n=0,1,...$) can then be orthonormalized
using the standard Gram-Schmidt method.

\section {Conclusions}

We have argued that if for a given representation the space of physical
states spanned by the wormhole wave functions can be equipped with a Hilbert
structure, that space must coincide with the Hilbert space of the Lorentzian
gravitational system under consideration. According to Ashtekar's proposal
[9,10], the inner product in that space can then be determined by imposing a
set of Lorentzian reality conditions. The wormhole states supply us in this
way with a basis for the physical Hilbert space of the Lorentzian quantum
theory. The elements of that basis are designated by a set of wormhole
parameters which, in order to be well-defined quantum numbers that
characterize the wormhole states, have to coincide with the eigenvalues
reached in the wormhole wave functions by a complete set of compatible
observables. In fact, the spectra of those observables must be equal to the
ranges of the wormhole parameters for the wormhole solutions to form a
complete set of physical states.

We have proved the correctness of the above statements
in the particular case of
a Friedmann-Robertson-Walker spacetime minimally coupled to a massless scalar
field. For this minisuperspace, the wormhole wave functions were already
known in the literature [5-7]. We have found the inner product selected by the
Lorentzian reality conditions, and shown that the Hilbert space of the model
admits a basis of normalizable wormhole states. We have considered in detail
two normalizable bases of wormholes, one of them given by an infinite but
countable number of wormhole solutions. The elements of any of these two
bases turn out to be eigenfunctions of a certain observable of the system,
and the corresponding eigenvalue provides the parameter that
describes the different wormholes. We notice that, in this case, the
eigenvalue of one observable is enough to characterize a quantum state,
because the minisuperspace possesses only one physical degree of freedom. On
the other hand, from the existence of a discrete basis of normalizable
wormholes, it follows that the Hilbert space of the model is separable and
admits an orthonormal basis.

Actually, a similar analysis to that presented here can also be carried out
in the two other minisuperspace models in which a complete set of wormhole
wave functions has already been obtained, i.e., a Friedmann-Robertson-Walker
spacetime conformally coupled to a massless scalar field [7] and a
Kantowski-Sachs spacetime provided with a minimally coupled scalar field with
no mass [8]. In those models, the space of wormholes can be supplied as well
with a Hilbert structure, and the corresponding Hilbert space can also
be seen to coincide with the Hilbert space of the Lorentzian quantum
theory. The explicit results for these minisuperspaces will be published
elsewhere [19].

If there always exists a basis of wormhole wave functions for the Hilbert
space of Lorentzian gravity, every quantum gravitational state can be
interpreted as a superposition of wormholes. The amplitudes of the
contributions of the different wormhole solutions in those superpositions
provide a complete description of the proper physical states of the theory.
On the other hand, if the quantum state of the Universe can be represented
as a normalizable wave function, such a wave function must then be given by a
particular combination of wormholes. This implies, in particular, that the
wave
function of the Universe must be damped for large three-geometries.
Whatever the requirements to determine the quantum state of the Universe
may be, they should guarantee this asymptotic behaviour.  One should then
expect the kind of boundary conditions usually employed in quantum cosmology
(i.e., the no-boundary proposal [20] and the ``tunneling'' proposals of
Linde [21] and of Vilenkin [22]) not to pick out, in general, a normalizable
wave function to represent the Universe, since these conditions do not
ensure the vanishing of the wave function for asymptotic large
three-geometries. Therefore, these boundary conditions must be either
modified or replaced in order to select a proper physical state that can
describe our Universe. Let us finally comment that, if the wormhole solutions
can be defined in full gravity as Euclidean path integrals [6], it must be
possible to obtain the wave function of the Universe as a particular sum
over histories on manifolds that possess an asymptotic region of large
three-volume where there are no gravitational excitations. In this sense,
the asymptotic flatness of our Universe would be a prediction of
the quantum theory of gravity.

{\bf Acknowledgements}

\vspace*{.4cm}
The author wants to thank P. Gonz\'alez D\'{\i}az, J. Mourao and C. Soo
for helpful conversations. This work was
supported by funds provided by the Spanish Ministry of Education and
Science Grant No. EX92-06996911.


\newpage

\begin{thebibliography}{28}

\bibitem{1} Hawking S W 1988 {\it Phys. Rev.} D {\bf 37} 904

\bibitem{2} Giddings S and Strominger A 1988 {\it Nucl. Phys.} B {\bf 306} 890

Hosoya A and Ogura W 1989 {\it Phys. Lett.} B {\bf 225} 117

Halliwell J J and Laflamme R 1989 {\it Class. Quantum Grav.} {\bf 6} 1839

\bibitem{3} Coleman S 1988 {\it Nucl. Phys.} B {\bf 310} 643

Hawking S W 1990 {\it Nucl. Phys.} B {\bf 335} 155

\bibitem{4} Bochner S 1946 {\it Bull. Am. Math. Soc.} {\bf 52} 776

Cheeger J and Gromoll D 1972 {\it Ann. Math.} {\bf 96} 413

\bibitem{5} Hawking S W and Page D N 1990 {\it Phys. Rev.} D {\bf 42} 2655

\bibitem{6} Garay L J 1991 {\it Phys. Rev.} D {\bf 44} 1059

\bibitem{7} Garay L J 1993 {\it Phys. Rev.} D {\bf 48} 1710

\bibitem{8} Campbell L M and Garay L J 1991 {\it Phys. Lett.} B
{\bf 254} 49

\bibitem{9} Ashtekar A 1991 {\it Lectures on Non-Perturbative Canonical
Gravity} ed Fang L Z and Ruffini R (Singapore: World Scientific)

\bibitem{10} Ashtekar A 1993 {\it Gravitation and Quantizations, Les Houches
Summer School Proceedings Vol. LVII} ed Julia B and Zinn-Justin J
(Amsterdam: North Holland)

\bibitem{11} Ashtekar A 1986 {\it Phys. Rev. Lett.} {\bf 57} 2244; 1987 {\it
Phys. Rev.} D {\bf 36} 1587

\bibitem{12} Rendall A 1993 {\it Class. Quantum Grav.} {\bf 10} 2261

\bibitem{13} Ashtekar A, Romano J D and Tate R S 1989 {\it Phys. Rev.} D {\bf
40} 2572

\bibitem{14} Mena Marug\'an G A 1993 Reality Conditions for Lorentzian and
Euclidean Gravity in the Ashtekar Formulation, {\it Preprint} Penn State
University CGPG-93/11-2, to appear in {\it Int. J. Mod. Phys.} D

\bibitem{15} Ashtekar A, Tate R and Uggla C 1993 {\it Int. J. Mod.
Phys.} D {\bf 2} 15

\bibitem{16} Mena Marug\'an G A 1993 Reality Conditions in Non-Perturbative
Quantum Cosmology, {\it Preprint} Penn State
University CGPG-93/9-3, to appear in {\it Class. Quantum Grav.}

\bibitem{17} Abramowitz M and Stegun I A (ed) 1970 {\it Handbook of
Mathematical Functions} 9th edn (Natl. Bur. Stand. Appl. Math. Ser. No. 55)
(Washington, D.C.: U.S. Govt. Print. Off.)

\bibitem{18} Gradshteyn I S and Ryzhik I M 1980 {\it Table of Integrals,
Series
and Products} 4th edn (San Diego: Academic Press)

\bibitem{19} Mena Marug\'an G A 1994 in preparation

\bibitem{20} Hawking S W 1982 {\it Astrophysical Cosmology} ed Br\"uck H A,
Coyne G V and Longair M S (Vatican City: Pontificia Academia Scientarium);
1984 {\it Nucl. Phys.} B {\bf 239} 257

Hartle J B and Hawking S W 1983 {\it Phys. Rev.} D {\bf 28} 2960

\bibitem{21} Linde A 1984 {\it Zh. Eksp. Teor. Fiz.} {\bf 87} 369 (1984 {\it
Sov. Phys. JETP} {\bf 60} 211); 1984 {\it Nuovo Cimento} {\bf 39} 401;
1984 {\it Rep. Prog. Phys.} {\bf 47} 925

\bibitem{22} Vilenkin A 1984 {\it Phys. Rev.} D {\bf 30} 509; 1986 {\it Phys.
Rev.} D {\bf 33} 3560; 1988 {\it Phys. Rev.} D {\bf 37} 888


\end{thebibliography}
\end{document}